# AN UPPER LIMIT TO THE MASS OF A CLOSE COMPANION CANDIDATE TO σ ORI E

*By José A. Caballero[1,2], Hervé Bouy[2] & Jorge Lillo-Box[3,2]*

[1]*Landessternwarte Königstuhl (ZAH), Heidelberg, Germany*
[2]*Centro de Astrobiología (CSIC-INTA), Madrid, Spain*
[3]*European Southern Observatory, Santiago de Chile, Chile*

The famous, very young, helium rich, magnetically-active, radio and X-ray emitter, short-period rotationally variable, spectroscopically peculiar star σ Ori E may have a close late-type stellar companion, which could explain flaring activity observed in some σ Ori E X-ray light curves. In 2009, Bouy *et al.* announced the detection of a faint companion candidate in the *Ks* band at 0.330 arcsec (130 AU) to the B2 Vp primary. Here, we carry out *z'*-band lucky imaging with *AstraLux* at the 2.2 m Calar Alto telescope in an attempt to constrain the properties of the companion candidate to σ Ori E. We impose a maximum mass of $2.0^{+0.2}_{-0.1}$ $M_{sol}$ and an earliest spectral type of K2±1, which leaves the door open to a new, inexpensive, near-infrared, adaptive-optics study.

*Introduction*

The helium-rich peculiarity of the spectrum of the B2 Vp star σ Ori E (48 Ori E, V1030 Ori, HD 37479, Mayrit 42062) was first noticed in 1956 by Berger[1], and two years later made widely public by Greenstein & Wallerstein[2]. Six decades later, dedicated works on σ Ori E have received over 1000 citations from approximately 50 dedicated works (*e.g., The magnetic field of* σ *Ori E*[3], *The rigidly rotating magnetosphere of* σ *Ori E*[4], *Shell and photosphere of* σ *Ori E – New observations and improved model*[5], *A new phenomenon in the spectrum of* σ *Ori E*[6], *etc.*). Actually, for decades the star attracted more attention than the rest of stars in the 3 Myr-old σ Orionis cluster, which is now a cornerstone for studies of circumstellar discs, coronal emission, accretion, and, especially, the mass function down to the substellar limit and beyond[7-9]. With an early B spectral type and a mass of 7-8 $M_{sol}$, σ Ori E is one of the most massive stars in the homonymous Trapezium-like system σ Ori, which also gives the name to the cluster and illuminates the Horsehead Nebula[10-12].

The B2 Vp star σ Ori E is abnormally rich in helium[13-17] and magnetically active[18,19]; it is indeed the prototype of magnetic peculiar Bp stars. Furthermore, it also displays periodic variability of 1.19 d in optical photometry and spectroscopy[17,20-25], photospheric and wind absorption lines[5,26,27], linear and circular polarization[3,18,28-30], and emission in Hα[6,31,32], non-thermal radio[33-37], and X-rays[38-41]. According to the most accepted scenarios proposed by Nakajima[42], Groote & Hunger[43], and Townsend *et al.*[4], the origin of the variability resides on abundance heterogeneities on the stellar surface that rotate rigidly with, and are connected through, radiatively driven wind streams to at least two clouds of confined plasma in a circumstellar magnetosphere. The very different inclinations of both stellar magnetic and rotational axes shape such a weird configuration.

Besides that, strong X-ray flares in σ Ori E were reported by Groote & Schmitt with *ROSAT*[44] and Sanz-Forcada *et al.* with *XMM-Newton*[45]. Observed flares are typical in young late-type stars, but virtually missing in early-type stars like σ Ori E. Magnetohydrodynamic simulations supported a centrifugal breakout hypothesis that could explain the reconnection heating and following X-ray flaring from the B2 Vp star[46,47]. However, centrifugal breakout was afterwards ruled out with an accurate photometric monitoring of σ Ori E with the *MOST* microsatellite[25]. Another alternative for explaning the X-ray flares, proposed by Caballero *et al.*[41,48], is the existence of an unresolved K-M-type stellar companion. In this scenario, σ Ori E would become a Lindroos binary system made of a B-type primary and a late-type companion[49-51].

In 2009, using multi-conjugate adaptive optics in the near infrared, Bouy *et al.*[52] discovered a companion candidate to σ Ori E, 3–4 mag fainter in *Ks*, at only $\rho \sim$ 0.330 arcsec and $\theta \sim$ 301 deg (note the corrected position angle[12]). In spite of the numerous works devoted to σ Ori E, this companion candidate at the limit of resolution and sensitivity of the Bouy *et al.* observations has never been confirmed. At the cluster distance of 385 pc[11,53], the measured angular separation translates into a projected physical separation of 130 AU. The radius of σ Ori E is of the order of 3–4 $R_{sol}$[25,30], while the circumstellar magnetosphere with its two plasma-confinement regions could extend to up to 2.4 stellar radii[30] (*i.e.*, about 0.04 AU). In spite of the apparently large physical separation between the companion candidate to σ Ori E and its circumstellar magnetosphere, the existence of wind collision and/or magnetic channeling might explain the high order variations to the rigidly rotating magnetosphere model[4,19]. If it were not the case, a (young, active) late-type companion may at least explain the observed X-ray flares observed by Groote & Schmitt and Sanz-Forcada *et al*.

In this work, we present new, red-optical, lucky-imaging observations of very high spatial resolution with the aim of characterizing the faint companion to σ Ori E and of improving the uncertain measures of position and flux by Bouy *et al.*

*Observations and analysis*

On 22 Oct 2014, we observed σ Ori E ($V$ = 6.61 mag) with the lucky-imaging camera *AstraLux Norte*[54] in service mode under director's discretionary time. *AstraLux* was attached to the Cassegrain focus of the 2.2 m telescope of the Calar Alto Observatory, in Almería, Spain. We got 50 000 frames of 15 ms in frame-transfer mode for each of the two used passbands, SDSS *i'* and *z'*. We fixed the gain for both filters, did windowing in a 256 × 256-pixel sub-array, used the right part of the electron-multiplying high-speed CCD for excluding the bad column #242, chose the quadrant less affected by dust, and set all other parameters to default. Total exposure time per filter was 750 s. For a precise astrometric calibration, we also observed a field centred on the M3 globular cluster.

We did a standard reduction of the data with the dedicated *AstraLux* pipeline[54], by selecting only the best 1.0, 2.5, and 5.0% of all frames, and by shifting, stacking, and re-sampling them into a nearly diffraction-limited image[55,56] (the *AstraLux* pipeline also provides the option of selecting the best 10% frames). The final plate scale was 0.02327 arcsec/pixel. Unfortunately, the *i'*-band image showed an apparent elongation in the north-south direction, probably due to incorrect focussing, and could not be used for the following analysis. The full final *z'*-band image, together with a *SofI*/NTT image with a larger field of view for comparison, is shown in Fig. 1.

Fig. 2 shows the largest detectable difference of magnitude between primary and secondary in our stacked *AstraLux z'*-band image as a function of angular separation to σ Ori E. It was computed from the average of the 3σ noise measurements in increasing concentric annuli of 1-pixel width centred on our target (the north-south read-out "spikes" barely affects the standard-deviation calculation). The derived maximum magnitude limit is valid for angular separations between 0.25 and 4.6 arcsec. The hunchbacked point-spread function, which avoids quantifying any standard full-width-half-maximum, is typical of all *AstraLux* images. At the largest separations to σ Ori E, the minimum magnitude difference of our stacked image is as high as $\Delta z'$ = 8.0 mag, which translates into an apparent magnitude depth of $z'$ = 15.0 mag (see below). This companion detectability method is widely used in the literature; see an example by the authors and with this instrument, including hunchbacked point-spread function, generation of artificial signals, and identification of false positives**,** in Lillo-Box *et al.*[56].

*Results and discussion*

The close companion candidate to σ Ori E is at 0.33 arcsec according to Bouy *et al*. At such angular separation, and with an uncertainty of 10%, we imposed a minimum magnitude difference $\Delta z'$ = 2.9±0.3 mag for the 2.5% best *AstraLux Norte* frames. The magnitude difference for the 1 and 5% best frames is identical within error bars.

We made a simple interpolation between well-calibrated magnitudes of σ Ori E in neighbouring passbands[57,58] for estimating its apparent magnitude in the *z'* band at 7.04±0.04 mag (σ Ori E saturates in Sloan Digital Sky Durvey DR9[59] *z'* band). With the *AstraLux* value of $\Delta z'$ > 2.9±0.3 mag, we concluded that the companion candidate must be fainter than $z'$ = 10.0±0.3 mag. At the σ Orionis distance[53] (distance modulus = $7.93^{+0.08}_{-0.15}$ mag), this limit translates into a minimum absolute magnitude $M_{z'}$ = 2.1±0.3 mag (see Table I). This magnitude is brighter than that of the most massive, hottest stars in the Lyon BT-Settl models[60] for 3 Myr (1.4 $M_{sol}$, 4700 K). Thus, we used the Siess *et al.*[61] models for 3 Myr, solar metallicity $Z$ = 0.02 and no overshooting for putting limits on the astrophysical parameters of the companion candidate.

After making another simple interpolation (the Siess *et al.* models do not tabulate absolute magnitudes in *z'*, but in *R*, *I*, *J* and *H*, just to mention the neighbouring passbands), the companion candidate must be less massive than $2.0^{+0.2}_{-0.1}$ $M_{sol}$, cooler than $T_{eff}$ = $4960^{+190}_{-70}$ K, and later than K2±1. At 3 Myr, while the B2 Vp primary is already in the main sequence, the companion is still in the contracting phase and has lower surface gravity (and greater mass) than field K dwarfs of the same effective temperature.

The depth of our stacked image allowed also us to discard the existence of lower-mass companions at larger separations. In our *AstraLux* image, we were not able to identify another source apart from σ Ori E; in their deeper *MAD* near-infrared images, Bouy *et al.* did not find any other source at less than 5 arcsec to σ Ori E either. In Table II, we show the upper limits of masses and spectral types at four angular separations. We used the Siess *et al.* models at 1.0 arcsec (as at 0.33 arcsec), and the BT-Settl models for 3 Myr and the $T_{eff}$–spectral type conversion of Reid & Hawley[62] for 2.0 and 3.0 arcsec.

The earliest spectral type of the companion candidate to σ Ori E derived from our *AstraLux* observations, K2±1, agrees with the "late-K to early-M" spectral-type interval hypothesised by Caballero *et al.*[48] from the existence of X-ray flares. The minimum magnitude difference in *z'* between primary and candidate companion is also consistent with the 3−4 mag difference in *Ks* estimated by Bouy *et al.* The ESA *Gaia* mission will unlikely be able to resolve any very close companion 7−8 mag fainter in *G* than the bright primary[63]. Besides, given the very low cluster proper motion (μ < 5 mas/yr) and its location in the antapex[64], an astrometric follow-up of the pair may not be feasible, even with *Gaia*. Therefore, new high-resolution multi-band imaging in the red optical and/or near-infrared, deeper by about 2 mag than our current *AstraLux* data, are necessary to constrain the properties of the companion candidate at 0.33 arcsec to σ Ori E. The *Hubble Space Telescope* could easily answer this dilemma (the primary star may be too bright for the *James Webb Space Telescope*), but a near-infrared adaptive-optics system at a 4−8 m ground telescope, such as *NACO*, *SPHERE*, *GPI* or *MagAO*, could do it faster and cheaper in just 10 min of observing time (around 30 min including overheads). The over 1000 citations and six decades of time-consuming works on σ Ori E may deserve this small observational effort.

Table I

*Apparent and absolute magnitudes in the red optical and near-infrared of σ Ori E and its close companion candidate*

| | *σ Ori E* | | *Companion candidate* | |
|---|---|---|---|---|
| *Band* | *Apparent magnitude [mag]* | *Absolute magnitude [mag]* | *Apparent magnitude [mag]* | *Absolute magnitude [mag]* |
| *R* | 6.84±0.01 | $-1.09^{+0.08}_{-0.15}$ | ... | ... |
| *I* | 7.08±0.01 | $-0.85^{+0.08}_{-0.15}$ | ... | ... |
| *z'* | 7.04±0.04 | $-0.89^{+0.10}_{-0.16}$ | >10.0±0.3 | >2.1±0.3 |
| *J* | 6.974±0.026 | $-0.96^{+0.08}_{-0.15}$ | ... | ... |
| *H* | 6.954±0.031 | $-0.98^{+0.09}_{-0.15}$ | ... | ... |

Table II

*Upper limits to the presence of resolved companions to σ Ori E*

| *ρ* [arcsec] | *s* [AU] | *z'* [mag] | $M_{max}$ [$M_{sol}$] | *Earliest sp. type* |
|---|---|---|---|---|
| 0.33 | 130 | 10.0±0.3 | $2.0^{+0.2}_{-0.1}$ | K2±1 |
| 1.00 | 385 | 10.9±0.3 | 0.65±0.15 | K4±1 |
| 2.00 | 770 | 12.6±0.3 | 0.69±0.09 | K7±1 |
| 3.00 | 1160 | 14.3±0.3 | 0.25±0.04 | M3±1 |

*Acknowledgements*

We thank the anonymous referee for his prompt and polite comments. Based on observations collected at the Centro Astronómico Hispano Alemán (CAHA) at Calar Alto, operated jointly by the Max–Planck-Institut für Astronomie and the Instituto de Astrofísica de Andalucía. Financial support was provided by the Spanish MICINN and MINECO under grants AYA2014-54348-C3-2-R, AYA2011-30147-C03-03, RYC-2009-04666 and RYC-2009-04497.

*References*

(1) J. Berger, *Contr. Inst. Ap. Paris*, **A**, No. 217, 1956.
(2) J. L. Greenstein & G. Wallerstein, *ApJ*, **127**, 237, 1958.
(3) J. D. Landstreet & E. F. Borra, *ApJ*, **224**, L5, 1978.
(4) R. H. D. Townsend, S. P. Owocki & D. Groote, *ApJ*, **630**, L81, 2005.
(5) D. Groote & K. Hunger, *A&A*, **116**, 64, 1982.
(6) N. Walborn, *ApJ*, 191, L95, 1974.
(7) F. Walter *et al.*, *Handbook of star forming regions, Volume I*, 2008, p. 732.
(8) J. A. Caballero, *A&A*, **478**, 667, 2008.
(9) K. Peña-Ramírez *et al.*, *ApJ*, **754**, 30, 2012.
(10) J. A. Caballero, *A&A*, **466**, 917, 2007.
(11) S. Simón-Díaz, J. A. Caballero, J. Lorenzo, *ApJ*, **742**, 55, 2011.
(12) J. A. Caballero, *The Observatory*, **134**, 273, 2014.
(13) C. Chadeau, *Annales d'Astrophysique*, **23**, 935, 1960.
(14) D. A. Klingesmith *et al.*, *ApJ*, **159**, 513, 1970.
(15) A. P. Odell, *ApJ*, **194**, 645, 1974.
(16) S. J. Adelman & D. M. Pyper, *A&AS*, **62**, 279, 1985.
(17) M. A. Smith & D. A. Bohlender, *A&A*, **475**, 1027, 2007.
(18) R. H. D. Townsend *et al., ApJ*, **714**, 318, 2010.
(19) M. E. Oksala *et al., MNRAS*, **419**, 959, 2012.
(20) B. Thomsen, *A&A*, **35**, 479, 1974.
(21) D. Groote & K. Hunger, *A&A*, **52**, 303, 1976.
(22) D. Groote & K. Hunger, *A&A*, **56**, 129, 1977.
(23) J. E. Hesser, P. P. Ugarte & H. Moreno, *ApJ*, **216**, L31, 1977.
(24) P. Harmanec, *Bulletin of the Astrophysical Institutes of Czechoslovakia*, **35**, 193, 1984.
(25) R. H. D. Townsend *et al., ApJ*, **769**, 33, 2013.
(26) H. Pedersen & B. Thomsen, *A&AS*, **30**, 11, 1977.
(27) S. N. Shore & D. N. Brown, *ApJ*, **365**, 665, 1990.
(28) J. C. Kemp & L. C. Herman, *ApJ*, **218**, 770, 1977.
(29) D. Clark & P. A. McGale, *A&A*, **205**, 207, 1988.
(30) A. C. Carciofi *et al., ApJ*, **766**, L9, 2013.
(31) C. T. Bolton, *ApJ*, **192**, L7, 1974.
(32) A. Reiners *et al., A&A*, **363**, 585, 2000.
(33) S. A. Drake *et al., ApJ*, **322**, 902, 1987.
(34) S. A. Drake *et al., ApJ*, **420**, 387, 1994.
(35) K. A. Miller, R. B. Philips & Ph. Andre, *AAS*, **181**, 1902, 1992.
(36) F. Leone & G. Umana, *A&A*, **268**, 667, 1993.
(37) P. Leto *et al.*, *MNRAS*, **423**, 1766, 2012.
(38) S. J. Wolk, PhD thesis, State University of New York at Stony Brook, 1996.
(39) S. L. Skinner *et al., ApJ*, **683**, 796, 2008.
(40) N. R. Hill *et al.*, *IAUS*, **272**, 194, 2011.
(41) J. A. Caballero, J. F. Albacete-Colombo & J. López-Santiago, *A&A*, **512**, A45, 2010.
(42) R. Nakajima, *Ap&SS*, **11**, 285, 1985.
(43) D. Groote & K. Hunger, *A&A*, **319**, 250, 1997.
(44) D. Groote & J. H. M. M. Schmitt, *A&A*, **418**, 235, 2004.
(45) J. Sanz-Forcada, E. Franciosini & R. Pallavicini, *A&A*, **421**, 715, 2004.

(46) R. H. D. Townsend & S. P. Owocki, *MNRAS*, **357**, 251, 2005.
(47) A. ud-Doula, R. H. D. Townsend, S. P. Owocki, *ApJ*, **640**, L191, 2006.
(48) J. A. Caballero *et al.*, *AJ*, **137**, 5012, 2009.
(49) K. P. Lindroos, *A&A*, **156**, 223, 1986.
(50) J. H. M. M. Schmitt *et al.*, *ApJ*, **402**, L13, 1993.
(51) N. Huélamo *et al.*, *A&A*, **359**, 227, 2000.
(52) H. Bouy *et al., A&A*, **493**, 931, 2009.
(53) J. A. Caballero, *MNRAS*, 383, 750, 2008.
(54) F. Hormuth *et al.*, *SPIE*, **7014**, E48, 2008.
(55) M. Janson *et al., ApJ*, **754**, 44, 2012.
(56) J. Lillo-Box, D. Barrado & H. Bouy, *A&A*, **566**, A103, 2014.
(57) J. R. Ducati *et al.*, *ApJ*, **558**, 309, 2001.
(58) M. F. Skrutskie *et al., AJ*, **131**, 1163, 2006.
(59) C. P. Ahn *et al., ApJS*, **211**, 17, 2014.
(60) I. Baraffe *et al.*, *A&A*, **577**, A42, 2015.
(61) L. Siess, E. Dufour & M. Forestini, *A&A*, **358**, 593, 2000.
(62) I. N. Reid & S. L. Hawley, *New Light on Dark Stars Red Dwarfs, Low-Mass Stars, Brown Stars*, 2005.
(63) J. H. J. de Bruijne, *Ap&SS*, **341**, 31, 2012.
(64) J. A. Caballero, A&A, **514**, A18, 2010.

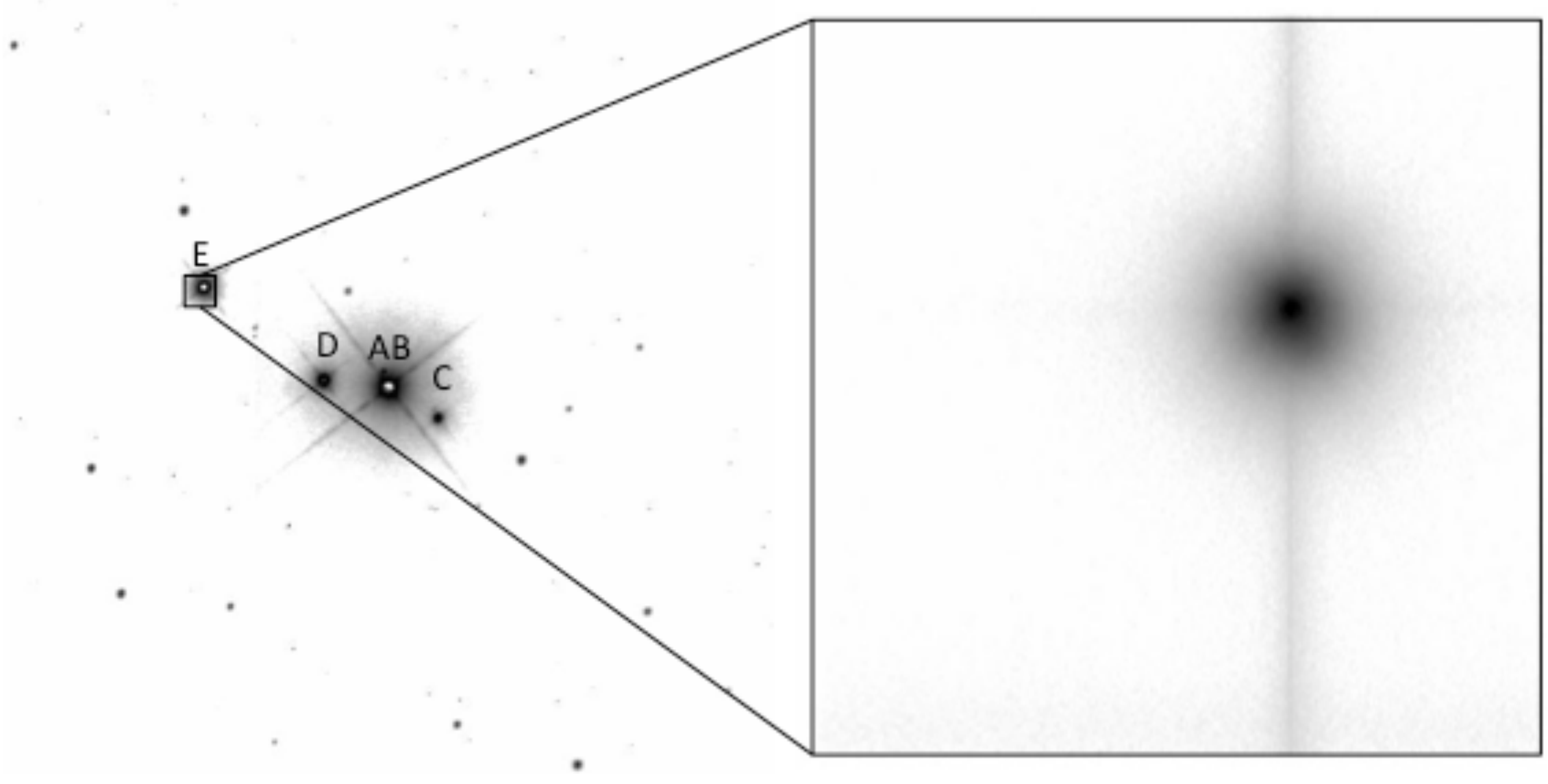

FIG. 1

*Left*: part of a *SofI*/3.6 m New Technology Telescope *J*-band natural-seeing image centred on σ Ori Aa,Ab,B; approximate size is 150 x 150 arcsec$^2$. The brightest stars of the Trapezium-like system are labelled. *Right*: *AstraLux Norte*/2.2 m Calar Alto *z'*-band lucky image roughly centred on σ Ori E; size is 6.0 × 6.0 arcsec$^2$. In both images, north is up and east is left. Note the very small size *AstraLux* field of view with respect to the *SofI* one.

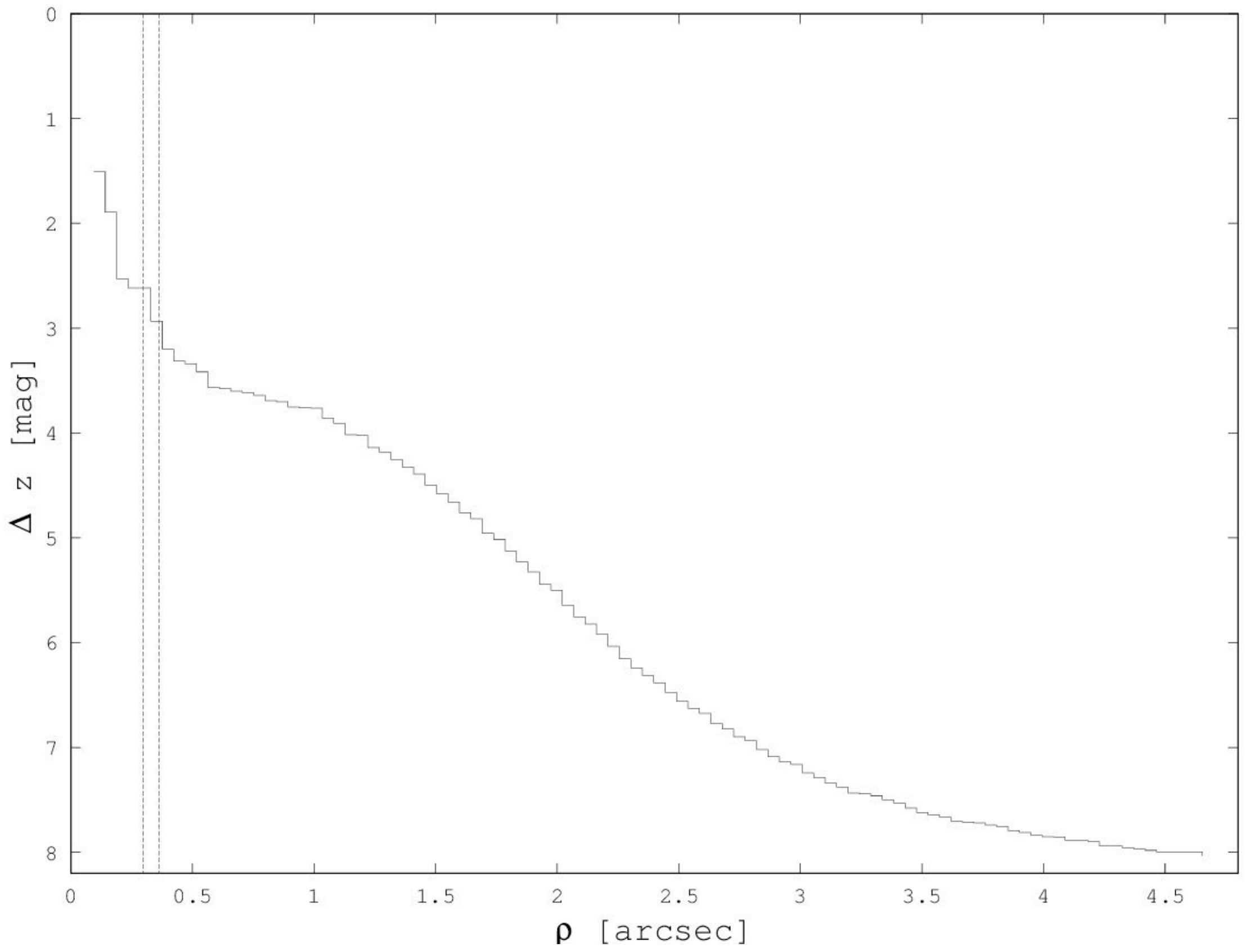

FIG. 2

Limit of our *AstraLux Norte* $z'$-band image: magnitude difference $\Delta z'$ as a function of angular separation $\rho$ to σ Ori E (2.5% best frames). The vertical dotted lines indicate the expected location of the candidate companion to σ Ori E, at $\rho$ = 0.33±0.03 arcsec.